\documentclass[usenatbib,usegraphicx]{mn2e}
\usepackage{aas_macros}
\usepackage{rotating}
\usepackage{longtable}
\usepackage{times} 
\def\arcs{\hbox{$^{\prime\prime}$}}
\def\arcm{\hbox{$^{\prime}$}}

\def\lsim{~\raise0.3ex\hbox{$<$}\kern-0.75em{\lower0.65ex\hbox{$\sim$}}~}
\def\gsim{~\raise0.3ex\hbox{$>$}\kern-0.75em{\lower0.65ex\hbox{$\sim$}}~}
\newcommand{\mum}{$\,\mu$m}

\newcommand{\hst}{\textsl{HST}}
\newcommand{\sirtf}{\textsl{SIRTF}}
\begin{document}
\title[The HDF-N SCUBA Super-map I]
{The HDF-North SCUBA Super-map I: Submillimetre maps, sources and number counts}

\author[Borys et. al.]{
\parbox[t]{\textwidth}{
\vspace{-1.0cm}
Colin Borys$^{1,2}$,
Scott Chapman$^{2},$
Mark Halpern$^{1}$,
Douglas Scott$^{1}$
}
\vspace*{6pt}\\
$^{1}$ Department of Physics \& Astronomy,University of British Columbia,
       Vancouver, BC, Canada \\
$^{2}$ California Institute of Technology, Pasadena, CA 91125, USA
\vspace*{-0.5cm}}

\date{Accepted 20 May 2003}

\maketitle

\begin{abstract}
We investigate the emission of sub-millimetre-wave radiation from galaxies in
a 165 square arcminute region surrounding the Hubble Deep
Field North.  The data were obtained from dedicated observing runs from our
group and others using the SCUBA camera on the James Clerk Maxwell Telescope,
and combined using techniques specifically developed  for low signal-to-noise
source recovery.  The resulting `Super-map' is derived from about 60 shifts
of JCMT time, taken in a variety of observing modes and chopping strategies,
and combined here for the first time.  At $850\,\mu$m we detect 19 sources
at ${>}\,4\sigma$, including 5 not previously reported.  We also list an
additional 15 sources between 3.5 and $4.0\sigma$ (where 2 are expected by
chance).  The $450\,\mu$m map contains 5 sources at ${>}\,4\sigma$.
We present a new estimate of the 850\mum\ and 450\mum\ source counts.
The number of sub-mm galaxies we detect
account for approximately 40\% of the 850\mum\ sub-mm background, and we 
show that mild extrapolations can reproduce it entirely.  A clustering
analysis fails to detect any significant signal in this sample of SCUBA 
detected objects.
A companion paper describes the multiwavelength properties of the sources.
\end{abstract}

\begin{keywords}
methods: statistical - methods: numerical -
large-scale structure of Universe - galaxies: formation
\vspace*{-1.25cm}
\end{keywords}

\section{Introduction}
The Hubble Deep Field North (HDF-N) \citep{1996AJ....112.1335W}, a small
region of the sky targeted by the \textsl{Hubble Space Telescope (HST)}, has
stimulated the study of the high redshift Universe ever since the data were
released in 1995.  The original optical image of the HDF is one of the deepest
ever obtained, and resolves thousands of galaxies in an area of a few square
arcminutes.  However, since optical images capture only a narrow part of the
spectrum, and since the rest-frame wavelength range detected depends on the
redshift, it is necessary to supplement the \hst\ image with data at other
wavelengths in order to obtain a more complete understanding of galaxy
evolution.  In the years since the HDF image became public, deep pointings
using radio, X-Ray, near-IR, and mid-IR telescopes have been conducted.
Additionally, optical spectroscopy has been carried out on hundreds of
suitable objects in the field, thereby obtaining redshifts for most of the
brighter \hst-detected objects.

The original HDF field is the size of a single WFPC2 field of view,
roughly $2\arcm\times2\arcm$.  \hst\ also obtained shallower observations in
fields adjacent to this, extending the region of study to roughly
$6\arcm\times6\arcm$.  These `flanking fields' have also been covered by other
telescopes, and in fact over time the region associated with the HDF has been
extended to about $10\arcm\times10\arcm$ in most wave-bands. For an excellent
review of the HDF-N region and its impact on the optical view of astronomy,
refer to the article by \citet{2000ARA&A..38..667F}.
Despite these extensive observations, fully understanding the high redshift
universe is hindered by the presence of dust, which re-processes radiation and
emits it in the Far-Infrared (FIR).  The importance of this is clearly
demonstrated in the population of galaxies detected by the Submillimetre Common
User Bolometer Array (SCUBA; Holland et al.~1999). 
Even with the tremendous observational effort of the past five years, 
we have made only modest progress with SCUBA detected galaxies beyond the conclusions 
drawn from the pioneering work of
\citet{1998Natur.394..241H} and \citet{1997ApJ...490L...5S}.  SCUBA mapping
surveys have constrained the number counts, and show that significant evolution is
required in the local Ultra-Luminous Infra-Red Galaxy (ULIRG) population in
order to explain the abundance of high redshift SCUBA sources.

Having detected these sources, the goal now is to characterise them and try to
answer fundamental questions about their nature.  What
powers their extreme luminosities?  What is their redshift distribution?
What objects do they correspond to today?
One way to do this is by doing pointed photometry on a list of high redshift
sources detected at other wavelengths, but this obviously introduces some
biases.  An alternate approach, one taken by many groups, is to do a
`blank field' survey, and compare the sub-mm map against images taken with
other telescopes at other wavelengths.  Arguably the best field to do this is
in the Hubble Deep Field region, which has more telescope time invested in
observations than any other extragalactic deep field.  In addition to the data
already available, observations with the \hst-ACS and  upcoming
confusion-limited \sirtf\ observations, make this an appealing region
to target with SCUBA.
For all these reasons it is worthwhile to combine the available sub-mm
data in this part of the sky.

\section{Observations}
Observations with SCUBA require the user to `chop' the secondary mirror at a rate of 7.8125 Hz between the target position and a reference (also called `off') position.  The difference between the signals measured at each position removes common-mode atmospheric noise.  The direction and size of the chop throw is adjustable, and is typically set such that the off position does not change over time due to sky rotation.  A map made in such a way will then exhibit a negative copy of the source in the off position.  

In raster-scan mode, the whole telescope moves on the sky in order to sample a region larger than the array size.  A map made from data collected in this mode is commonly referred to as a `scan-map'.  An alternative way to sample a large area is to piece together smaller `jiggle' maps.  In this mode, the telescope stares toward the target and the secondary mirror steps around a dither pattern designed to fill in the space on the image plane between the conical feedhorns.  Photometry mode is very similar to jiggle-mapping, except the dither pattern is smaller in order to spend more time observing the target object.  The trade off is that the image plane is not fully sampled, and there will be gaps in a map made from photometry data.  Jiggle-mapping and photometry also differ from scan mapping because they use two off positions in the chopping process.  Therefore there will be two negative echos of each source.

Our group has produced a shallow `scan-map' of the region at 450 and 850\mum\
(Borys et al.~2002, hereafter Bo02) using SCUBA.
Given the high profile of the region however, groups from the UK and Hawaii
also targeted the HDF to exploit the rich multi-wavelength observations
available. The $\sim 3\arcm\times3\arcm$ area centred on the HDF itself,
studied originally by Hughes et al.~(1998; hereafter H98), was recently
re-analysed by Serjeant at al.~(2002; hereafter S02) and we use their
published data for comparison here.  Results from observations taken by the
Hawaii group can be found in Barger et al.~(2000; hereafter Ba00).
We have obtained all of these data from the JCMT archive or from the observers
directly and embedded them in our scan-map.   The extra data considerably
increase the sensitivity of the final map in the overlap regions, but at the
expense of much more inhomogeneous noise and a correspondingly more
complicated data analysis.  Co-adding data taken in different observing modes
has not previously been performed for extra-galactic sub-mm surveys, and so we
discuss our procedure in some detail below.

The full list of projects allocated time to study the HDF region is extensive.
Observing details for each project are given in Table \ref{hdftab}. Almost all
work carried out in the region has used the jiggle-map mode in deep, yet
small surveys. Three projects involved targeted photometry observations taken
in `2-bolometer chopping' mode.  The SURF User Reduction Facility
({\textsc SURF}; Jenness \& Lightfoot 1998) software was not able to process
these particular files for inclusion in the map, but we shall discuss these
observations later.  Project M00BC01 is a single jiggle-map observation
taken in such a way that co-adding it to the map is also difficult, and
shall be described later as well.  All of the remaining data could be co-added
into a `Super-map', as we now describe.

\begin{table*}
\caption[Summary of SCUBA HDF region observations.]{Summary of SCUBA HDF region
observations.  Type refers to either photometry (P), jiggle-map (J), or
scan-map (S).  Approximate 1$\sigma$ noise estimates in mJy are provided,
though they do vary somewhat across the individual images.  The area surveyed
in square arcminutes is also listed. Projects denoted by an asterisk were
photometry observations that could not be added directly to the map, due to
the inability of {\textsc SURF} to extract the bolometer positions for
two-bolometer chopping observations. In addition, M00BC01 was taken using an
unfortunate chopping configuration which cancelled out 2 sources near the 
western edge of the map. These data are not folded directly into the map.
In total, there have been almost 60 shifts (of 8 hours each) in which the
telescope was observing this region, which is considerably more than the
original HST optical imaging.}

\label{hdftab}
\begin{tabular}{lllllr}
\\\hline
Project ID & Type & Shifts & $\sigma_{450}$ (mJy) & $\sigma_{850}$ (mJy) & Area (arcmin$^2$)  \\ \hline\hline
M97BU65        & P,J    & 17 &  4  & 0.4 & 9    \\
M98AC37        & P      &  2 & 65  & 2   & 3    \\
M98AH06        & P      &  4 & 11  & 1   & 3    \\
M98BC30        & S      &  2 & 75  & 6   & 160  \\
M98BU61        & P      &  1 & 33  & 3   & 2    \\
M99AC29        & S      &  2 & 95  & 6   & 160  \\
M99AH05        & J      & 11 & 20  & 2   & 39   \\
M99BC41$^*$    & P      &  1 & N/A & N/A & N/A  \\
M99BC42        & P,J,S  &  3 &120  & 6   & 160  \\
M00AC16$^*$    & P      &  1 & N/A & N/A & N/A  \\
M00AC21$^*$    & P      &  1 & N/A & N/A & N/A  \\
M00AH07        & J      &  4 & 40  & 2   & 12   \\
M00BC01        & J      &  1 & N/A & 3   &  5   \\
M01AH31        & J      &  7 & 25  & 2   & 22   \\  
\hline
\end{tabular}
\end{table*}

\section{Data Reduction}
\begin{figure*}
\begin{center}
\includegraphics[width=6in,angle=0]{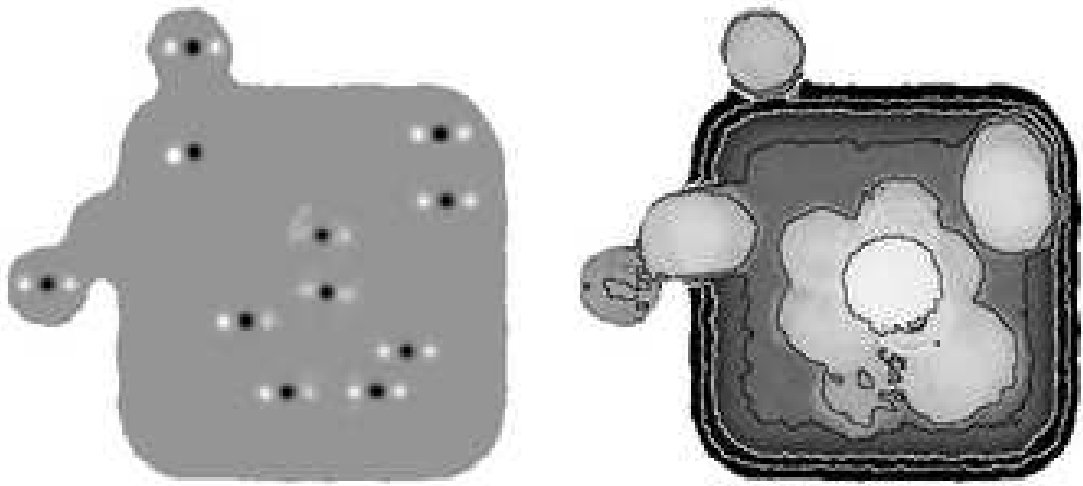}
\caption[Layout of the SCUBA HDF observations.]{Layout of the SCUBA HDF
observations.  The majority of the data is contained within a
$12\arcm\times12\arcm$ square, but other jiggle-maps were also taken
within the wider {\sl Chandra\/} or VLA images.  The left panel is a map of
the HDF using simulated sources but the real pointing information. This allows
the variable PSF to be easily seen.  Black here corresponds to sources, and
white to the negative echoes caused by the necessity to chop. Various parts of
the field were observed in different ways, as is evident from the different
PSFs.  On the right is the actual noise map at 850\mum\ with contours at
1, 3, 5, 7, and 9\,mJy overlaid.  The deepest pointings of
\citet{1998Natur.394..241H} are in the centre of the image, while other
jiggle-maps and photometry observations can be seen superimposed on our
scan-map.  The corresponding images for the 450\mum\ map are very similar in
appearance, though with higher noise levels.}
\label{fig:hdflayout}
\end{center}
\end{figure*}

We start by simply regridding the bolometer data onto an output map, ignoring
for a moment the effect of a strongly varying effective point-spread function
(PSF, i.e. the effect of the different chops) across the field.
Fig.~\ref{fig:hdflayout} demonstrates just how different the observations are.
Shown is the PSF at various locations in the field, alongside the 850\mum\
noise map with contours overlaid.  Because the telescope is moving at a rate
of 3 arcsec per sample while taking scan-map data, there is no advantage to
choosing a pixel size smaller than 3 arcsec.  Some other studies have chosen
smaller (1 arcsec) pixels for jiggle-maps, but given the uncertainties in
JCMT pointings, we feel that a choice of 3 arcsecond pixels is sufficient.

\subsection{Flux and pointing calibration}
Registering the individual data-sets with respect to each other is difficult,
due to the lack of bright sources in the field.  The pointing log for each
night of data was inspected, and in all but a few cases there is no reason to
distrust the pointing; pointing checks were always performed on the same side
of the meridian as the HDF, and observations were not conducted through
transit\footnote{During much of the period in which this data was obtained, the JCMT telescope would drop slightly in elevation while tracking an object through transit}.   Although individual night-to-night pointing comparisons between
data sets is not feasible, it is still possible to check for a systematic
pointing offset between scan- and jiggle-maps.  One might be concerned about
errors on the order of 3 arcsec, due to the speed at which the telescope moves
while scanning.  

In a map made solely from the scanning observations, a group of sources is
detected that correspond to the same objects seen in the deep jiggle map
in Ba00 (project M99AH05).  A least squares comparison between a
$4\arcm \times 4\arcm$ square in the scan-map against the same area in the
jiggle-map indicates a $4\pm4$ arcsecond offset along the direction of the
chop (110 degrees east of north).  

Exploiting the well established FIR/radio correlation, we also performed a pointing check using the following procedure:  We took a list of $\mu$Jy VLA 1.4\,GHz sources detected by \citep{2000ApJ...533..611R} and extracted the 850\mum\ flux from the super-map at each postions.  The average of these fluxes was used as a figure of merit in checking the pointing.  By shifting the super-map in both directions and calculating this `stacked' flux, we find that no significant positional offset is required maximize the average sub-mm flux at the radio positions.

For each night of jiggle-map and photometry observations, flux calibrations
were performed on suitable targets in the area. In each case these calibrations
were consistent with the averages determined for others taken in the same time
periods, and with the larger sample reported in \citet{2002MNRAS.336....1A}. Since individual calibrations may have larger statistical variation,
the average calibration value over the run was used instead.

Flux calibration of the scan-maps is more problematic.  Since this mode is
currently not so well characterised, it is important to compare the map
against those taken from the better understood jiggle-maps. Although pointing
seems well constrained, the fluxes in the HDF scan-map are generally larger
than their jiggle-map counterparts when using the `standard' gains.  This was
an issue first discussed in Bo02.  Unfortunately no usable scan-map calibration
measurements were taken during the observations.  Therefore we
proceed by adjusting the overall calibration of the scan-map and finding,
in a chi-squared sense, the value that minimizes the difference between the
two maps:
\begin{equation}
\label{equ:crossflux}
\chi^2(r) = \sum (S_{\rm Jiggle} - rS_{\rm Scan})^2.
\end{equation}
Here the sum is over all the pixels.  Because the beam patterns (two- versus
three-beam) are different, we use only those pixels within a beam-width of the
sources detected in the jiggle-maps.  Based on this analysis we determine that scan-map flux conversion factors are 0.8 times that of the standard jiggle-map calibrations.  

We obtained scan-map calibration data for other (unrelated to HDF) projects that observed planets, and estimated the flux conversion factor from them.  This analysis also obtains the factor of 0.8 difference between scan- and jiggle-map calibrations.

\subsection{Source detection}
No effort was made to deconvolve this combined map; the observing strategies
employed by both our group and others do not interconnect pixels very well,
and therefore a robust deconvolution cannot be performed.  However, in order
to gain back the extra sensitivity from the off-beams, we would like to fit
each pixel in the map to the multi-beam PSF.  This procedure has been adopted
by other groups as well, but the complication in this case is the variable PSF
across the field.  

Instead of fitting to the PSF, we can fold in the flux from the off-beams in a
time-wise manner.  For each sample, we add the measured flux to the pixel
being pointed to in addition to its negative flux (appropriately weighted) at
the position of the off-beam.  This is equivalent to performing one iteration
of the map-making procedure described in Johnstone et al. (2000; see also
Borys~2002).  A single Gaussian is then used to fit for sources in the final
map. This requires an image relatively free of sources that might lie in the
location of the off-beam of another source, but produces the same output as
one would obtain from fitting with the beam pattern.  The resulting image
should be considered not so much as a map but rather the answer to the
question: what is the best estimate of the flux of an isolated point source
at the position of each pixel?  The image can only be properly interpreted
in conjunction with the accompanying noise map.

\section{Monte-Carlo simulations}
\begin{figure}
\begin{center}
\includegraphics[width=3in,angle=0]{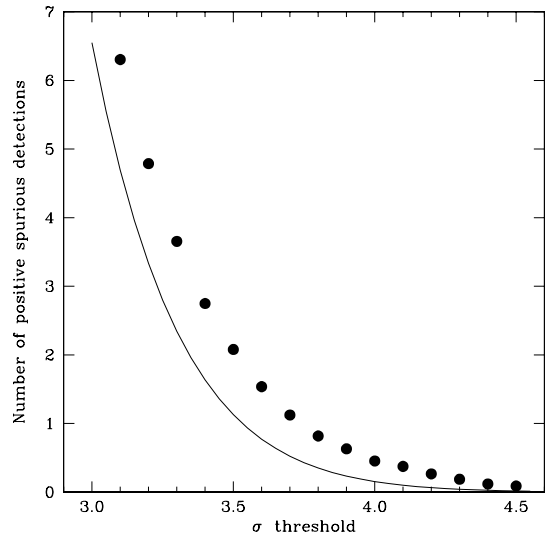}
\caption[Number of positive sources in the 850\mum\ map expected at random.]{Number of
positive sources in the 850\mum\ map expected at random. The filled circles show how
many spurious sources were detected in simulated maps made from noise alone.
Error bars are omitted for clarity, but are small because of the large number
of sources generated in the many simulations.  The number of positive sources one would expect by chance given Gaussian statistics and a
Gaussian-shaped beam with a FWHM of 14\arcs.7 in an 165 square arcminute region is plotted as the thick line.}
\label{fig:nbyrandom}
\end{center}
\end{figure}

There are several simulations that must be performed to assess the reliability
of the maps.  To determine how many detections might be false positives, we
created a map by replacing the 850\mum\ data with Gaussian random noise with a
variance equivalent to the noise estimated for each bolometer. The map was
then run through the same source-finder algorithm as the real data.  This was
repeated 100 times.  The average number of positive and negative detections as
a function of signal-to-noise threshold is plotted in Fig.~\ref{fig:nbyrandom},
along with the number one would expect based simply on Gaussian statistics and
the number of independent beam-sizes in the map (which is an underestimate
because it assumes well-behaved noise).  A $3.5\,\sigma$ cut is commonly used,
but these results suggest that at this threshold about four detections will
be spurious in our map (two positive and two negative detections).  Given that
the slope of this plot is still rather steep at $3.5\,\sigma$, small errors in
the noise estimate can lead to more false positives.  Therefore we adopt a
$4.0\,\sigma$ cut to determine sources in the HDF super-map.  Note that the
number of false positives for the 450\mum\ map will be four times larger
because the beam is half as big (of course if the noise is not well behaved
this can be even worse!).  Therefore, one wants to set a high detection
threshold for 450\mum\ objects (but as we will find later, there is only one
$>5\sigma$ source at 450\mum).  The situation is further complicated by the fact that the true background is not blank, and is popoulated by many unresolved sources that create confusion noise.

\begin{figure}
\begin{center}
\includegraphics[width=3.5in,angle=0]{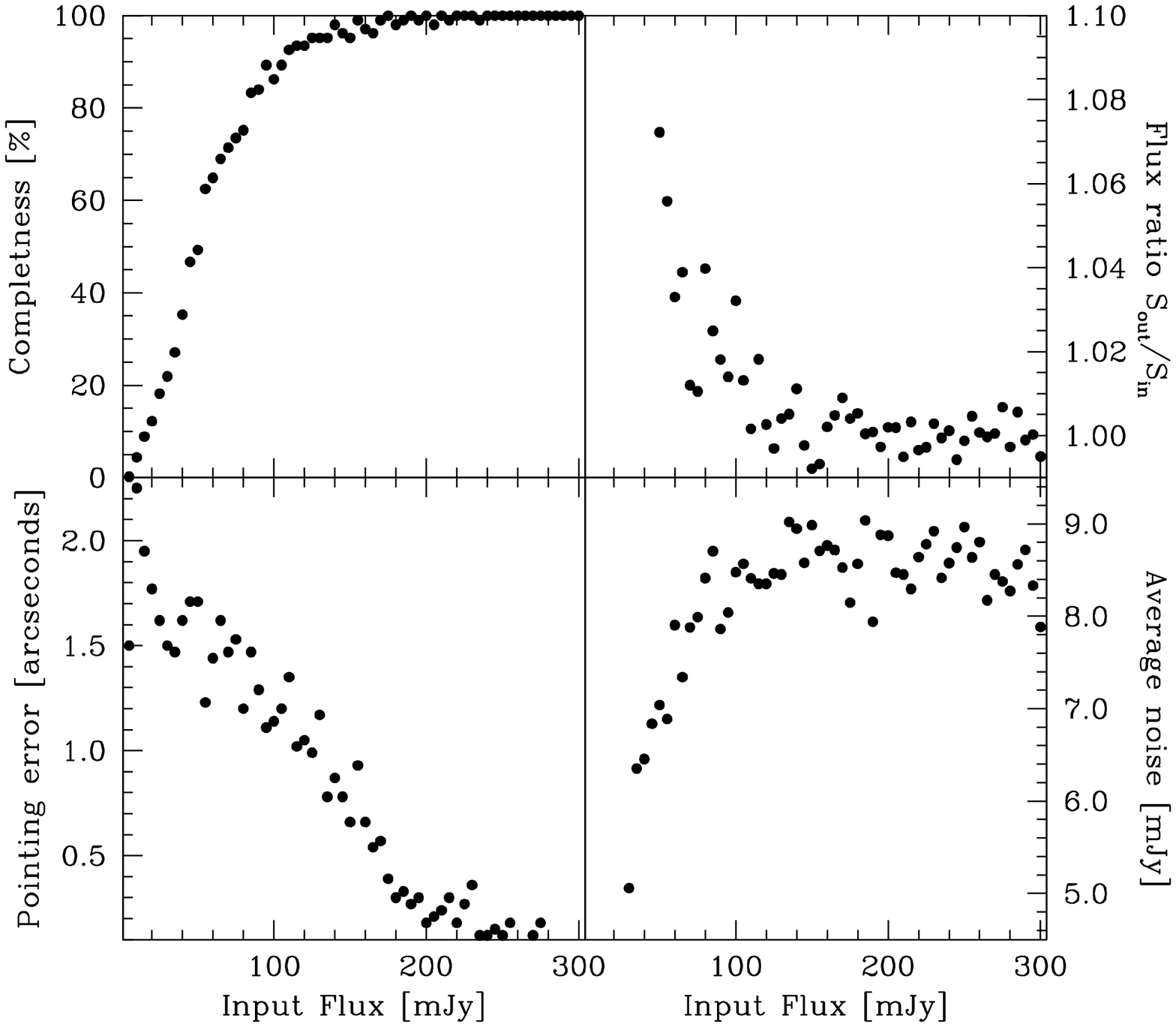}
\caption[450\mum\ source recovery Monte-Carlo results.]{450\mum\ source
recovery Monte-Carlo results.  From left to right and top to bottom the plots
are: a) completeness; b) flux bias ratio; c) pointing error; and d) noise
level. Error bars are not plotted for clarity, but one can gauge the 
uncertainty by the scatter of the points in the vertical direction.}
\label{fig:complete450}
\end{center}
\end{figure}

\begin{figure}
\begin{center}
\includegraphics[width=3.5in,angle=0]{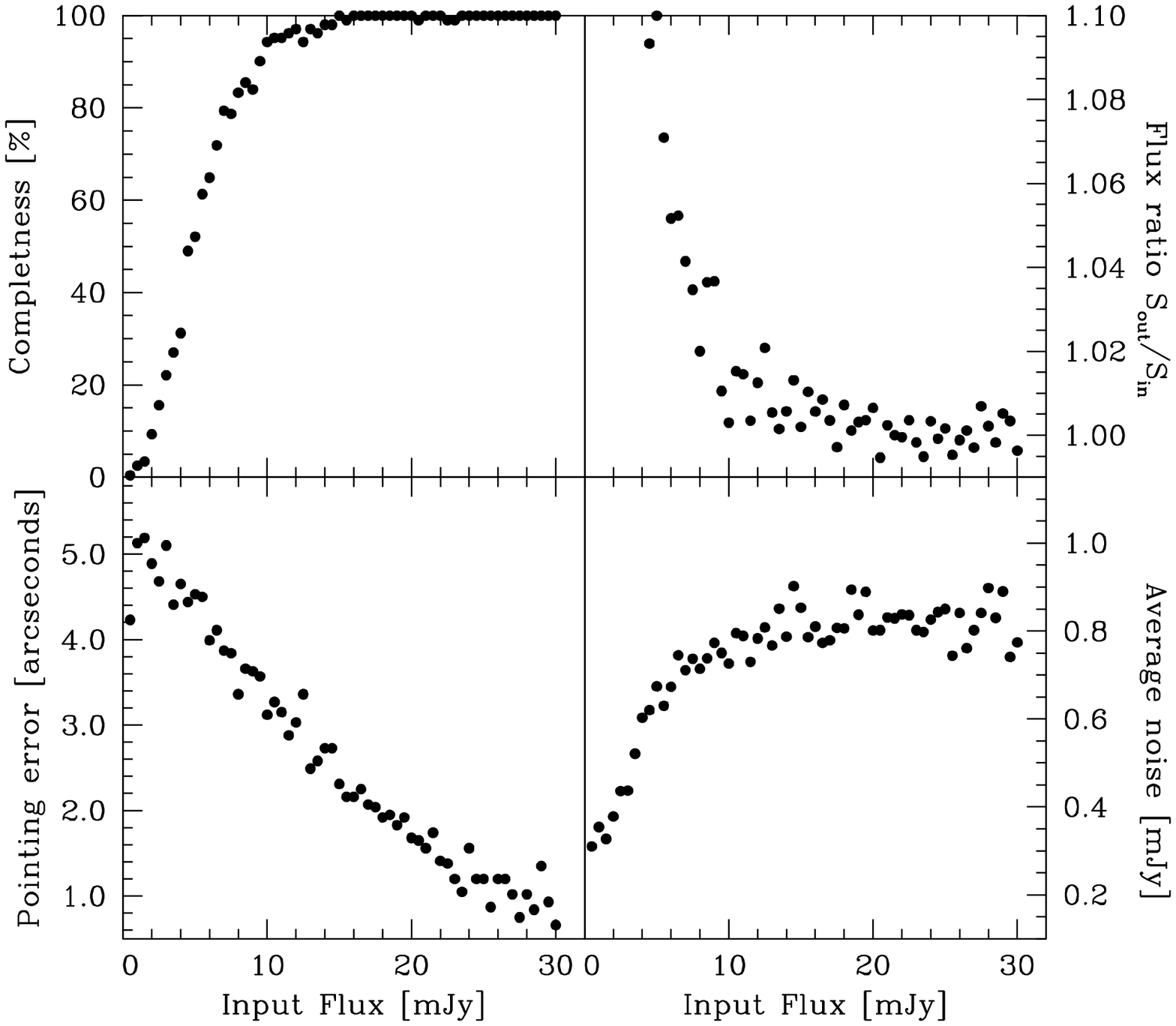}
\caption[850\mum\ source recovery Monte-Carlo results.]{850\mum\ source
recovery Monte-Carlo results. The order is the same as in the last figure.
In general the plots are very similar to their 450\mum\ counterparts, with
fluxes an order of magnitude smaller.
}
\label{fig:complete850}
\end{center}
\end{figure}

To further estimate the reliability of our detections, we added 500 sources
(one at a time) of known flux for a range of flux levels into the map and
attempted to extract them using the same pipeline as for the real data.  A
source was considered `recovered' if it was detected with a SNR greater than
4.0 and its position was within half the FWHM of the input position.  At each input flux, the position offset, flux bias, and noise of all the
recovered sources were averaged and plotted in
Figs.~\ref{fig:complete450} and \ref{fig:complete850} for the 450\mum\ and
850\mum\ maps, respectively.  

The panel showing completeness is self explanatory; it plots the percentage of
sources recovered in the Monte-Carlos as a function of input flux. As one
expects, it is difficult to recover faint sources around the noise limit, but
very bright sources are always recovered. We will discuss this issue more when
deriving source-counts.  

The ratio of output and input flux in the adjacent panel shows that sources 
fainter than the threshold have been scattered up due to the presence of 
noise and are `detected' with higher than their true flux.  The relevant
noise components in this case are instrumental noise and confusion noise.  
The latter is called Eddington bias \citep{1940MNRAS.100..354E}, and
dominates the central part of the map where the instrumental noise is smallest.
At 850\mum, the confusion limit is $\sim 1\,$mJy, and thus sources with 
instrumental noises around this level will be subject to Eddington bias.

The RMS of the difference
between input and output positions allows us to estimate a positional error for
real detections as a function of flux level.  As one would expect, the
uncertainty is smaller as the input flux goes up.  The final plot in the
sequence shows the average noise level associated with the recovered sources
as a function of input flux.  At the faint end, the noise is lower because
such sources are only detected in the deepest parts of the map.  As the flux
of the source increases above the noise level of the least sensitive region
(in this case the underlying scan-map), the average noise of the detections
levels off to the average noise level of the field.  The 450\mum\ plots
(Fig.~3) are very similar to those at 850\mum\ (Fig.~4), scaled by
approximately a factor of 10 in flux.

Basic conclusions at 850\mum\ are that at a flux limit of 8\,mJy the source
counts are about 80\% complete, fluxes are biased by only a few percent, and
positions are accurate to about 3.5 arcsec. The brighter objects are
constrained much better than this, but there are few of them.  At the
faintest levels confusion has a significant affect on fluxes, positions, and
completeness.

\section{Sub-mm sources in the HDF}
Most of the 850\mum\ sources exhibit off-beam signatures that are
distinguishable by eye (the negative echos to the left and right of the
sources).  Finding sources at 450\mum, however is more difficult.  The single
beam pattern is not well described by a Gaussian, plus the sensitivity at
450\mum\ is too poor to detect any but the brightest sources. In addition,
being more weather-dependent, the noise is even more inhomogeneous than for the
850\mum\ data.  Nevertheless, we will report the $3\sigma$ upper limit to the
450\mum\ flux for each 850\mum\ detection.  To avoid reporting spurious
detections, we have set a threshold of 4.0$\sigma$ on the 850\mum\ catalogue
derived from the super-map.  However, a supplementary list of sources at
850\mum\ detected above $3.5\sigma$ is also provided for comparison against
other data sets.  The full list of 850\mum\ and 450\mum\ sources is presented
in Table~\ref{tab:detectlist}.

\begin{table*}
\caption[Sub-mm detections in the HDF super-map.]{Sub-mm detections in the HDF
super-map. Only sources with a SNR (shown in brackets) greater than 3.5  are
listed.  If the source was previously detected in a separate survey, then the
reference to the paper is given, as well as the measured flux from there. The
papers are  Ba00:\citep{2000AJ....119.2092B}, Bo02:\citep{2002MNRAS.330L..63B},
S02:\citep{serjeant}. $3\sigma$ upper limits in the other channel are given
for each source.  }
\label{tab:detectlist}
\begin{tabular}{llllll}
\hline
ID &RA &DEC &$S_{850}\pm\sigma_{850}$ &$S_{450}\pm\sigma_{450}$ &Previously Detected\\\hline\hline
\multicolumn{6}{c}{850\mum\ detections $\geq4\sigma$}\\\hline
SMMJ123607+621145 &12:36:07.3 &62:11:45 &$15.2\pm3.8\,$(4.0$\sigma$)    &$<97   $& Bo02: $15.4\pm3.4$\,mJy\\
SMMJ123608+621251 &12:36:08.5 &62:12:51 &$16.0\pm3.7\,$(4.3$\sigma$)    &$<95   $& Bo02: $13.8\pm3.3$\,mJy\\
SMMJ123616+621518 &12:36:16.6 &62:15:18 &$6.3\pm0.9\,$(6.5$\sigma$)     &$<35   $&\\ 
SMMJ123618+621009 &12:36:18.9 &62:10:09 &$6.6\pm1.5\,$(4.3$\sigma$)     &$<66   $&\\
SMMJ123618+621554 &12:36:18.7 &62:15:54 &$7.2\pm0.9\,$(8.1$\sigma$)     &$<34   $& Ba00: $7.8\pm1.6$\,mJy\\
SMMJ123621+621254 &12:36:21.8 &62:12:54 &$12.1\pm2.6\,$(4.7$\sigma$)    &$<85   $& Bo02: $11.4\pm2.8$\,mJy\\
SMMJ123621+621712 &12:36:21.3 &62:17:12 &$8.8\pm1.5\,$(6.0$\sigma$)     &$<72   $& Ba00: $7.5\pm2.3$\,mJy \\
                  &           &         &                               &        & Bo02: $13.2\pm2.9$\,mJy\\
SMMJ123622+621618 &12:36:22.6 &62:16:18 &$8.6\pm1.0\,$(8.5$\sigma$)     &$<45   $& Ba00: $7.1\pm1.7$\,mJy\\
SMMJ123634+621409 &12:36:34.2 &62:14:09 &$11.2\pm1.6\,$(7.1$\sigma$)    &$<67   $& Ba00: $19.0\pm2.8$\,mJy\\
SMMJ123637+621157 &12:36:37.3 &62:11:57 &$7.0\pm0.8\,$(8.3$\sigma$)     &$<46   $&\\
SMMJ123646+621451 &12:36:46.3 &62:14:51 &$8.5\pm1.3\,$(6.4$\sigma$)     &$<47   $& Ba00: $10.7\pm2.1$\,mJy\\
                  &           &         &                               &        & Bo02: $11.4\pm2.9$\,mJy\\
SMMJ123650+621318 &12:36:50.6 &62:13:18 &$2.0\pm0.4\,$(5.3$\sigma$)     &$<11   $& S02: HDF850.4/5 $2.1\pm0.3$\,mJy\\
SMMJ123652+621227 &12:36:52.3 &62:12:27 &$5.9\pm0.3\,$(18.0$\sigma$)    &$<12   $& S02: HDF850.1 $5.6\pm0.4$\,mJy\\
SMMJ123656+621203 &12:36:56.6 &62:12:03 &$3.7\pm0.4\,$(8.8$\sigma$)     &$<16   $& S02: HDF850.2 $3.5\pm0.5$\,mJy\\
SMMJ123700+620912 &12:37:00.4 &62:09:12 &$8.6\pm2.1\,$(4.1$\sigma$)     &$<85   $& Ba00: $11.9\pm3.0$\,mJy\\
SMMJ123701+621148 &12:37:01.3 &62:11:48 &$4.0\pm0.8\,$(5.2$\sigma$)     &$<27   $& S02: HDF850.6 $6.4\pm1.7$\,mJy\\
SMMJ123703+621303 &12:37:03.0 &62:13:03 &$3.4\pm0.6\,$(5.3$\sigma$)     &$<25   $&\\
SMMJ123707+621412 &12:37:07.7 &62:14:12 &$9.9\pm2.5\,$(4.0$\sigma$)     &$<85   $&\\
SMMJ123713+621206 &12:37:13.3 &62:12:06 &$6.1\pm1.4\,$(4.3$\sigma$)     &$<40   $& Ba00: $8.8\pm2.0$\,mJy\\\hline
\multicolumn{6}{c}{Additional 850\mum\ detections $\geq3.5\sigma$ and $<4.0\sigma$}\\\hline
SMMJ123607+621021 &12:36:07.3 &62:10:21 &$13.5\pm3.7\,$(3.7$\sigma$)    &$<97   $&\\
SMMJ123608+621433 &12:36:08.5 &62:14:33 &$6.1\pm1.7\,$(3.6$\sigma$)     &$<60$  &\\
SMMJ123611+621215 &12:36:11.9 &62:12:15 &$12.8\pm3.4\,$(3.7$\sigma$)    &$<90$  & Bo02: $12.2\pm3.0$\,mJy\\
SMMJ123628+621048 &12:36:28.7 &62:10:48 &$4.4\pm1.2\,$(3.7$\sigma$)     &$<55$  &\\
SMMJ123635+621239 &12:36:35.6 &62:12:39 &$3.0\pm0.8\,$(3.6$\sigma$)     &$<41   $& S02: HDF850.7 $5.5\pm1.5$\,mJy\\
SMMJ123636+620700 &12:36:36.9 &62:07:00 &$22.1\pm5.6\,$(3.9$\sigma$)    &$<155$ &\\
SMMJ123648+621842 &12:36:48.0 &62:18:42 &$19.5\pm5.4\,$(3.6$\sigma$)    &$<154$ &\\
SMMJ123652+621354 &12:36:52.7 &62:13:54 &$1.8\pm0.4\,$(3.9$\sigma$)     &$<16$  & S02: HDF850.8 $1.7\pm0.5$\,mJy\\
SMMJ123653+621121 &12:36:53.1 &62:11:21 &$2.8\pm0.8\,$(3.6$\sigma$)     &$<40$  &\\
SMMJ123659+621454 &12:36:59.1 &62:14:54 &$5.2\pm1.4\,$(3.8$\sigma$)     &$<72$  &\\
SMMJ123706+621851 &12:37:06.9 &62:18:51 &$21.6\pm5.8\,$(3.8$\sigma$)    &$<178$ &\\
SMMJ123719+621109 &12:37:19.7 &62:11:09 &$7.2\pm2.0\,$(3.6$\sigma$)     &$<55$  &\\
SMMJ123730+621057 &12:37:30.4 &62:10:57 &$13.3\pm3.6\,$(3.7$\sigma$)    &$<98$  & Bo02: $14.3\pm3.2$\,mJy\\
SMMJ123731+621857 &12:37:31.0 &62:18:57 &$27.1\pm7.6\,$(3.6$\sigma$)    &$<286$ &\\
SMMJ123741+621227 &12:37:41.6 &62:12:27 &$23.7\pm6.1\,$(3.9$\sigma$)    &$<185$ &\\\hline
\multicolumn{6}{c}{450\mum\ detections $>4\sigma$}\\\hline
SMMJ123619+621127 &12:36:19.3 &62:11:27 &$<5.8$                 &$110\pm26\,$(4.2$\sigma$)   & \\
SMMJ123632+621542 &12:36:32.9 &62:15:42 &$<5.9$                 &$105\pm25\,$(4.2$\sigma$)   & \\
SMMJ123702+621009 &12:37:02.6 &62:10:09 &$<5.2$                 &$120\pm27\,$(4.4$\sigma$)   & \\       
SMMJ123727+621042 &12:37:27.4 &62:10:42 &$<10.4$                &$220\pm42\,$(5.2$\sigma$)   & \\       
SMMJ123743+621609 &12:37:43.8 &62:16:09 &$<24.0$                &$300\pm72\,$(4.2$\sigma$)   & \\\hline
\end{tabular}
\end{table*}

\begin{figure*}
\begin{center}
\includegraphics[width=6in,angle=0]{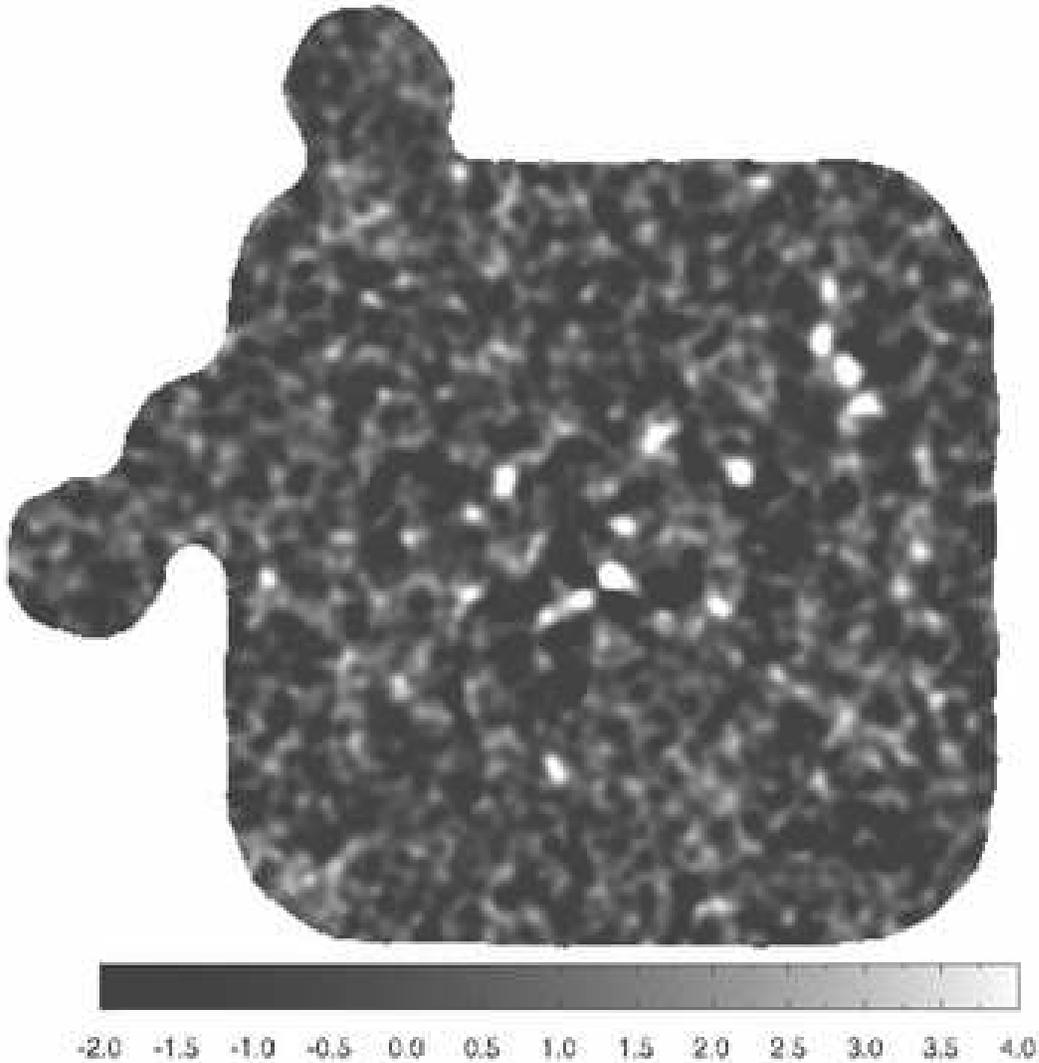}
\caption[The 850\mum\ signal-to-noise map.]{The 850\mum\ signal-to-noise map.
The greyscale is stretched to highlight sources.  This image can be used in
conjunction with the finder chart in Fig.~\ref{fig:finder} to identify the
objects reported in the source list.
}
\label{fig:hdfsmm850}
\end{center}
\end{figure*}

\begin{figure*}
\begin{center}
\includegraphics[width=6in,angle=0]{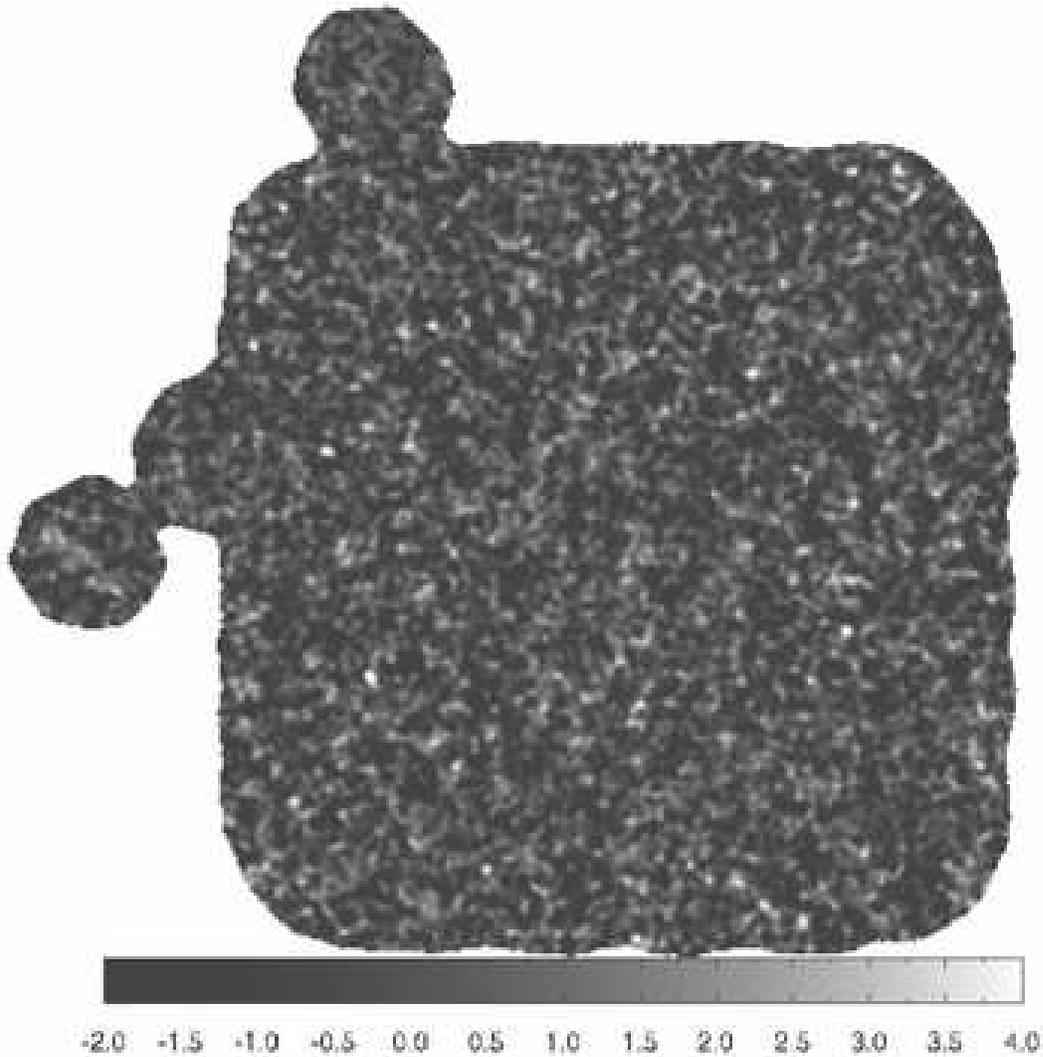}
\caption[The 450\mum\ signal-to-noise map.]{The 450\mum\ signal-to-noise map.
The greyscale is stretched to highlight sources.
}
\label{fig:hdfsmm450}
\end{center}
\end{figure*}

\begin{figure*}
\begin{center}
\includegraphics[width=5.0in,angle=0]{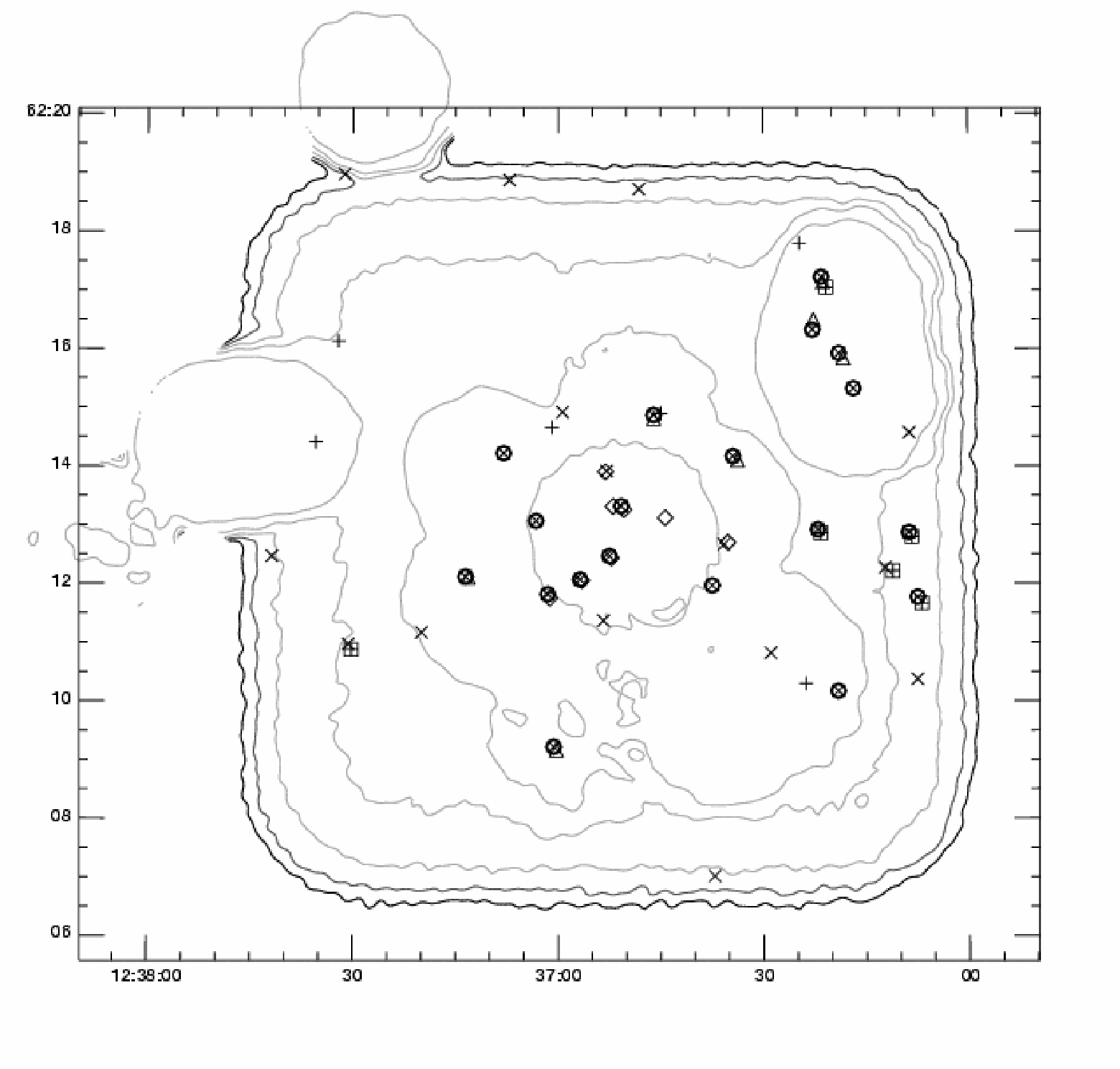}
\caption[Finder chart for sub-mm sources detected in the HDF super-map.]{Finder
chart for sub-mm sources detected in the HDF super-map.  The outlines are
850\mum\ noise contour levels from 1 to $11\,$mJy (darkest) in steps of
$2\,$mJy. Triangles are 850\mum\ source positions from
\citet{2000AJ....119.2092B}, diamonds from \citet{serjeant}, squares with an
embedded plus are the $4\sigma$ sources from \citet{2002MNRAS.330L..63B}, and
the plus symbols are the lower SNR sources from that survey.  Circles with
crosses are sources recovered above  $4\sigma$ from the super-map, and crosses
show those additional sources detected at lower significance.  The five
450\mum\ sources detected in the super-map are denoted with asterisks. 
}
\label{fig:finder}
\end{center}
\end{figure*}

A simple test of source reality is to search for negative flux objects.  All
but two negative sources found were associated with the off-beam of one of our
positive detections.  Based on the Monte-Carlos, this is not unreasonable.
There are 19 sources at 850\mum\ detected over $4\sigma$, 5 of which are new,
not having been reported before in the individual surveys.  Apart from a few
exceptional cases, described below, all sources previously reported in the
region are recovered at comparable flux levels. An additional 15 sources are
detected between $3.5\sigma$ and $4\sigma$.  Our Monte-Carlos suggest that on
average 2 of these may be spurious.  An image of the 850\mum\ super-map is
given in Fig.~\ref{fig:hdfsmm850}.  There are only 5 sources recovered from
the 450\mum\ image, which is shown in Fig.~\ref{fig:hdfsmm450}. There have
been no 450\mum\ detections previously reported in the HDF.  The finder chart
in Fig.~\ref{fig:finder} can be used to identify the sources extracted by our
algorithm in each of the maps.

\subsection{Comparing the source list against previous surveys}
\begin{figure}
\begin{center}
\includegraphics[width=3in,angle=0]{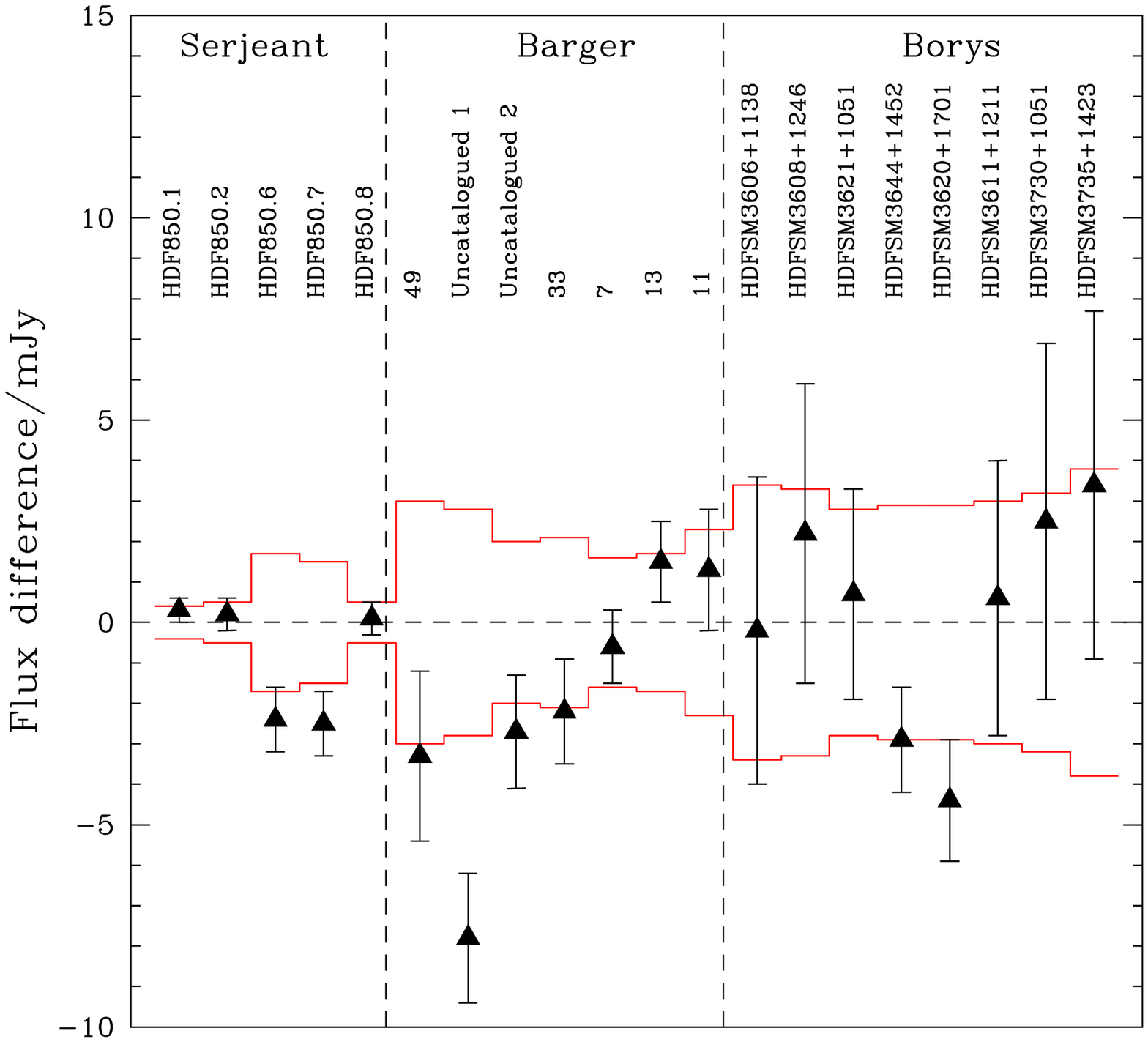}
\caption[Our new source list compared with previously reported detections.]{Our
new source list compared with previously reported detections.  We show the
difference between previously reported flux and that derived here for each
source in common.  The outline denotes the size of the error-bars from the
previously reported estimates. We subtract the reported fluxes from our own,
and plot the difference and error-bar as the filled triangles.   See the text
for a more complete description for those few points that seem discordant.
}
\label{fig:smfluxcmp}
\end{center}
\end{figure}

A plot comparing the recovered fluxes against those previously published is
given in Fig.~\ref{fig:smfluxcmp}.  We show the difference between the
published fluxes and super-map estimates. The flux differences should be zero,
and based on the size of the error bars it is clear that no significant
variations exist between the estimates.  The objects that appear slightly
discrepant are discussed at length below, and all sources will be described in
detail in a forthcoming paper. It should be noted that the error bars on the
sources recovered from the super-map are generally smaller than those obtained
from the individual sub-maps.

\subsubsection{The central HDF region from Hughes/Serjeant et al.}
Eight sources, labeled HDF850.1 through HDF850.8 are associated with the data
collected by H98.  The more thorough analysis in S02 found that one of them
(HDF850.3) detected in the original map fails to meet new detection criteria.
It is also the only source undetected in the super-map presented here.
HDF850.4 and HDF850.5 appeared to be a blend of two sources in their original
map, and therefore both H98 and SO2 took the extra step of attempting to fit
the amplitude and position of them. With the super-map however, this pair of
sources is better fit by a single source with flux comparable to the sum of
fluxes from the two extracted by S02. This may be because we use larger pixels
(3\arcs\ compared to 1\arcs\ from S02) and therefore lose some of the
resolution required to separate adjacent objects.  Sources HDF850.6 and
HDF850.7 are detected at a fainter level than reported in S02.  These
discrepancies might be partly due to their positions -- both are in a noisier
region of the individual sub-map, and near its edge.  The super-map contains
additional data (especially photometry) in the central HDF region, and so one
would expect our noise estimates to be mildly lower compared to those of S02, 
which is indeed generally true.

\subsubsection{HDF flanking field jiggle-maps from Barger et al.}
The seven sources detected individually in Ba00 are confirmed in the super-map
at comparable flux levels except for one object not detected in the radio.  This is
most likely due to differences in analysis methods; Ba00 used aperture
photometry with annuli centred on radio positions believed to be associated
with the sub-mm detection.  The source in question had no counterpart, and
therefore determining flux with aperture photometry is not as straightforward.
Even when the sub-map alone is considered individually, the measured flux does
not match that originally reported.

\subsubsection{Scan-map observations of Borys et al.}
All six sources from the $4\sigma$ list of Bo02 are recovered, although two of
them are detected at lower significance in the super-map. Four of the six
sources from the supplementary list of $3.5\sigma$ sources are not recovered
in the super-map.  Three of these exist in regions of overlap between surveys.
This supports the warning made in Bo02 that $4\sigma$ SCUBA detections are 
less likely to be spurious than lower SNR ones.
In general, however, we find that sources detected at ${>}\,4\sigma$ are
confirmed in separate sub-maps.  This is an important test of the reliability
of faint SCUBA detections, which can only be effectively carried out in the HDF
region, because of the overlapping independent data-sets.

\subsection{Comparison with photometry observations}
As we have mentioned, a number of photometry observations taken in the
two-bolometer chopping mode were conducted.  Although these data sets cannot
be co-added into the map, we can compare flux estimates for the photometry
bolometer with the position in the super-map.

\begin{figure}
\begin{center}
\includegraphics[width=3in,angle=0]{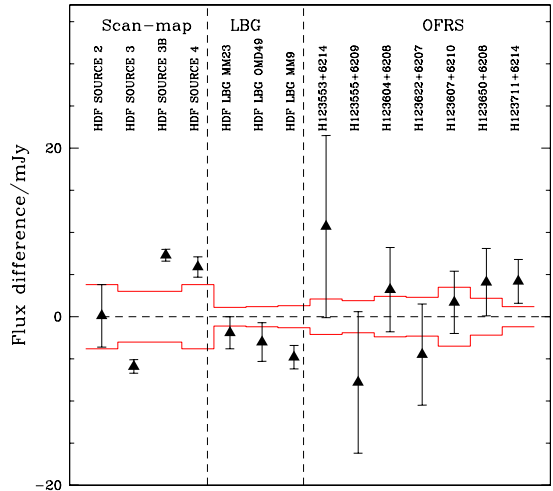}
\caption[Comparing super-map fluxes with photometry estimates.]{Comparing
super-map fluxes with photometry estimates.  We plot the difference between
reported photometry fluxes from observations not added into the map and those
derived from the super-map.  As before, the outline shows the error-bar from
the reported values, while our points (after subtracting off the reported
fluxes) are plotted as solid triangles.  Since some of the photometry goes
much deeper than our map, their error estimates are in those cases much lower.
All sources are in reasonable agreement within the combined error-bar of each
estimate.  Here, `LBG' is a Lyman-Break Galaxy target and `OFRS' is an
optically faint radio source.}
\label{fig:photomcomp}
\end{center}
\end{figure}

We should also note that while the scan-map observations were being taken,
several photometry observations were conducted at positions of tentative
detections in order to verify them.  None of these positions correspond to
detections in the final map, but all have fluxes consistent with the
corresponding position in the super-map.  This illustrates that the practice
of picking out low SNR sources `by eye' in SCUBA surveys is not very effective.  
Our group has also conducted two observing programmes designed to understand
the sub-mm properties of Lyman break galaxies (LBGs, Chapman et al.~2000)
and optically faint radio sources (OFRS, Chapman et al.~2002).  These types of
source will be discussed further in paper II.  Three measured LBGs that fall
within the region of the HDF are not detected in either the photometry
measurements or the super-map.  Of the seven OFRS photometry measurements, all
have comparable fluxes at their corresponding position in the super-map. A
summary plot similar to the one shown previously when comparing known
detections is given in Fig.~\ref{fig:photomcomp}.  Again, no significant
discrepancies exist between the estimates.

\section{Number counts of sub-mm sources}
In order to estimate the density of sources brighter than some flux threshold
$S$, $N({>}S)$ from our list of detected sources in Table~\ref{tab:detectlist},
we must account for several anticipated statistical effects:
\begin{enumerate}
\item The threshold for source detection, $S_{\rm T} = m\sigma$ is not uniform
because the noise varies dramatically across the map. 
\item Due to confusion and detector noise, sources dimmer than $S_{\rm T}$
might be scattered above the detection threshold and claimed as detected.
\item Similarly, sources brighter than $S_{\rm T}$ might be missed because of
edge effects, possible source overlaps, and confusion.
\end{enumerate}

Item~(iii) is simply the completeness of our list of sources. For a source
density $N(S)dS$ which falls with increasing flux, the effect of item (ii) can
exceed that from item (iii), resulting in an Eddington bias in the estimated
source counts.  One can calculate the ratio of the integrated source count to
the number of sources detected using the `detectability':
\begin{equation}
\label{equ:complete}
   {{1}\over{\gamma(S^\prime)}}= {{{\int_{S^\prime}^{\infty}
    {N(S)dS}}}\over{{\int_{0}^{\infty} \phi(S,S^\prime)N(S)dS}}}\,,
\end{equation}
where $N(S)dS$ is the number of sources with a flux between $S$ and $S+dS$.
We have introduced the quantity $\phi(S,S^\prime)$ which is the fraction of
sources between a flux, $S$ and $S+dS$, that are detected above a threshold,
$S^\prime$. Note that  $\phi(S,S^\prime)$ ranges between zero and unity, but
$1/\gamma(S^\prime)$ can be larger than one, depending on the form of $N(S)$
and choice of $S^\prime$. For bright sources, where the completeness is 1.0
and the SNR is high, $\phi(S,S^\prime)$ should approach unity.

The numerator of equation~\ref{equ:complete} is the quantity we are trying to
determine, and the denominator is the output from the survey.  To determine
$N(>{S})$ we simply take the raw counts from our survey and multiply them by
$\gamma(S^\prime)$.  
The calculation of $\gamma(S^\prime)$ from the Monte-Carlo estimates of
$\phi(S,S^\prime)$ requires a model of the source counts, of which there are
many different forms in the literature.  All share the property that they are
steep at the bright end and shallow at the faint end.  They are also typically
constrained such that the total amount of energy does not exceed the measured
value of the sub-mm extra-galactic background.  We employ the double-power law
form described in \citet{2002MNRAS.331..817S},
\begin{equation}
   {{dN(>S)}\over{dS}} = {{N_0}\over{S_0}} \left[{{\left({S}\over{S_0}\right)}^{\alpha}}
    + {{\left({S}\over{S_0}\right)}^{\beta}}\right]^{-1},
\end{equation}

and use $S_0=1.8\,$mJy, $\alpha=1.0$, $\beta=3.3$, and
$N_0=1.5\times10^4\,{\rm deg}^{-2}$. 

Our Monte-Carlos give us $\phi(S,S^\prime)$, which we fit using the form 
\begin{equation}
\phi(S,S^\prime)= 1- \exp\left(A(S-B)^C\right).
\end{equation}
for each value of $S^\prime$. This was chosen empirically to match the shape
of $\phi(S,S^\prime)$.  An alternative approach would be to perform more
Monte-Carlos with a finer spacing in $\Delta{S}$ and then spline the result to
allow for interpolation between the sample points -- the results are
almost identical.  Using these estimates of $\phi(S,S^\prime)$, together
with the source-count model, $\gamma(S^\prime)$ can be computed numerically
using equation~\ref{equ:complete}.  

The quantity $\gamma$ relates what we want to determine,  $N({>}S)$, with the
number of sources our survey detects. Obviously $\gamma$ is influenced by what
model is used in the calculation described above, and other forms of the
source spectrum are found in the literature.  We find $\gamma$ varies by no
more than 15\% across a wide range of reasonable parameter values.  This is
smaller than the Poissonian error caused by having so few sources detected.  

\subsection{The 850\mum\ number counts}
\begin{figure}
\begin{center}
\includegraphics[width=3.0in,angle=0]{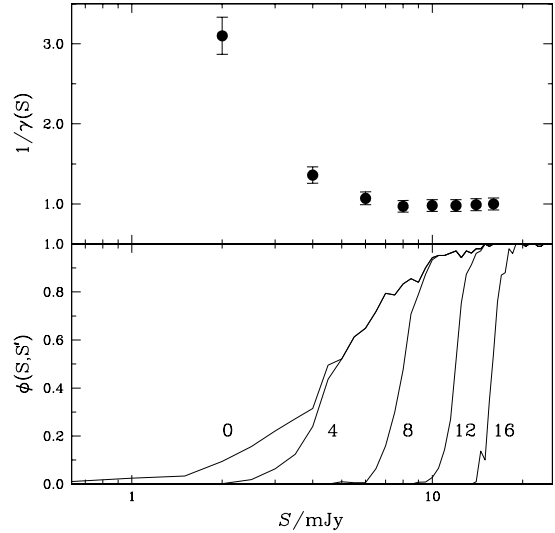}
\caption[Summary of the 850\mum\ source-count calculation.]{Summary of the
850\mum\ source-count calculation.  In the top panel we plot the detectability,
determined using the source-count model described in the text.  As expected,
it is higher at low flux levels (indicating that the survey has missed sources)
and approaches unity at the bright end.  Between 7 and 12\,mJy the
detectability is slightly lower than one, demonstrating that some sources
below the threshold are contributing.  The error bars represent the scatter
obtained when using a range of source-count models that roughly fit the
measured counts and sub-mm background.  In the lower panel, we plot
$\phi(S,S^\prime)$ for $S^\prime$ ranging from 0--16\,mJy in steps of 4\,mJy.
The  $S^\prime = 0$ case is simply the completeness of the survey.
}
\label{fig:nsmodel}
\end{center}
\end{figure}

In Fig.~\ref{fig:nsmodel} we plot the $\gamma$ determined using the source
count parameters described above, as well as $\phi(S,S^\prime)$ determined
from the Monte-Carlos.

\begin{figure*}
\begin{center}
\includegraphics[width=6in,angle=0]{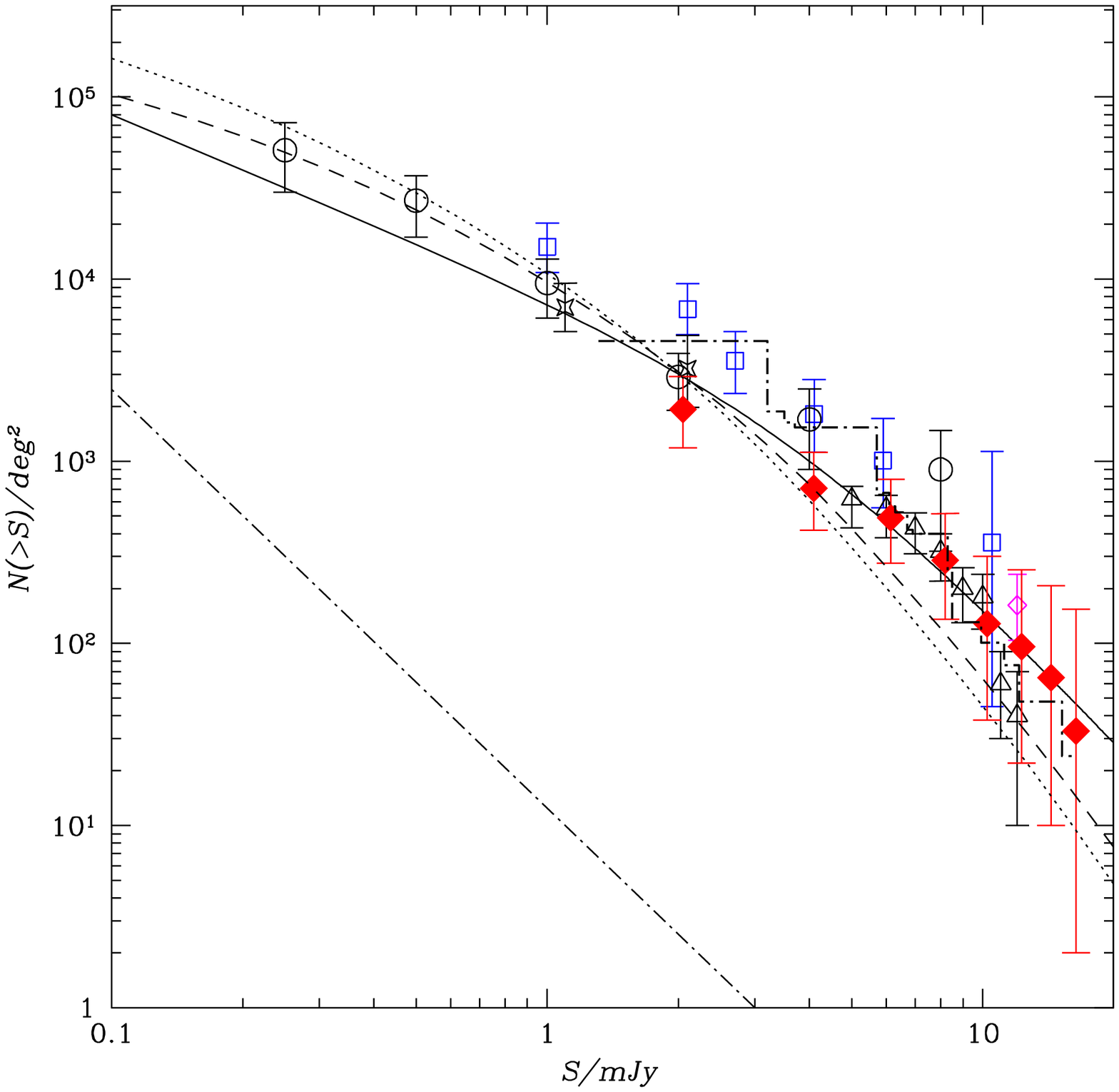}
\caption[The 850\mum\ cumulative source counts.]{The 850\mum\ cumulative
source counts.  The solid diamonds denote the results from the current work.
The step-function shows counts derived from the same data, but using a
different recipe (see Section~6.1). Counts derived from cluster studies by
\citet{2002MNRAS.330...92C} and \citet{2002MNRAS.331..495S} are shown by the
open squares and circles  respectively. The UK\,8-mJy survey counts
\citet{2002MNRAS.331..817S} are shown as open triangles.  Stars represent the
counts from \citet{1998Natur.394..241H}, and the open diamond is from
\citet{2002MNRAS.330L..63B}.  Some points are slightly offset along the flux
axis for clarity. Overlaid is the two power-law model used in the calculation
of $\gamma$ (solid line), and  two predictions based on representative galaxy
evolution models from \citet{2001ApJ...549..745R} -- the dashed line represents
a universe with $\Omega_{\rm M} = 1.0$ and $\Omega_\Lambda = 0.0$ while the
dotted line is $\Omega_{\rm M} = 0.3$ and $\Omega_\Lambda = 0.7$.  The
dot-dashed line below the curves is the count prediction obtained from
extrapolating the \textsl{IRAS} 60\mum\ counts and invoking no evolution.}
\label{fig:sourcecounts}
\end{center}
\end{figure*}

The detectability passes through unity at around 7\,mJy;  sources fainter than
this are not detected very efficiently, and therefore we must boost the raw
count to account for the incompleteness.  Past this, brighter source counts
get a slight boost from fainter sources that have been scattered above the
threshold due to noise.  Given $\gamma(S)$, we can now calculate the counts
based on the number of sources detected in the entire 165 square arcminute
super-map.  This is given in tabular form (see Table~\ref{tab:hdfsc}) and as a
plot alongside other estimates of the 850\mum\ source counts (in
Fig.~\ref{fig:sourcecounts}).

For the above source count approach we have provided estimates at a number
of flux values (chosen to be integer numbers of mJy).  There are other ways
of presenting the counts which avoid the need for binning.  An alternative
method uses the flux bias ratio (e.g.~a smooth curve fitted to the bottom
right panel of Fig.~4) to transform the flux of each detected source, and
then estimate the effective area in which such a source could have been
detected.  The results of this approach are shown by the step function plotted
in Fig.~\ref{fig:sourcecounts}, and are clearly consistent with the other estimate.

In general the derived counts are in excellent agreement with estimates from
other surveys.  They also agree broadly with the input model used to estimate
$\gamma(S)$. It should be noted that while other surveys use 68\% confidence
bounds, we prefer to quote 95\% limits.  It is interesting to note that the
bright counts from this survey and the UK~8--mJy project are slightly lower
than those found from surveys of smaller areas which often involve the
modeling of cluster lenses
\citep{2002AJ....123.2197C,2002MNRAS.331..495S,2002MNRAS.330...92C}.
Comparing the ratio of counts derived from our survey relative to the 8--mJy
project yields a factor of $\sim0.85$ for the 6--12 mJy range.  If we fit to
the average of the cluster counts and interpolate across this same range, this
ratio between our counts and theirs is roughly 0.4.  This might indicate that
those surveys had been sensitive to clustering, which would naturally lead to
an overestimate of the number counts.  It could also be that the cluster lens
surveys were contaminated by sources intrinsic to the clusters, or that there
was some systematic bias in the lens models.  We also note that some of the
surveys include candidates detected with a significance lower than the
$4\sigma$ cut we used here.

Although the current survey is limited by the lack of bright sources in the
HDF, it is clear that the counts fall off quite steeply with increasing flux.
Different surveys are often compared via the slope of a power-law form of the
number counts ($N(S)dS \propto S^{-\alpha}$).  For our counts we obtain
$\alpha=2.9\pm0.5$ (95\% confidence limits), which is in excellent agreement
with estimates obtain by other groups: $2.8\pm0.7$ from
\citet{1999ApJ...512L..87B}, $3.2_{-0.6}^{+0.7}$ from
\citet{1999ApJ...518L...5B}, and $3.2\pm0.7$ from \citet{2000AJ....120.2244E}.

At the faint end, it seems that the counts turn over and flatten out with
decreasing flux. Indeed if they do not then the sub-mm background will be
overproduced. Deeper surveys that reach to lower flux levels will be needed
in order to determine this unambiguously.

\begin{table*}
\caption[850\mum\ source counts from the HDF super-map using only the
$\ge4\sigma$ detections.]{850\mum\ source counts from the HDF super-map using
only the $\ge4\sigma$ detections. The total area surveyed is $165\,$arcmin$^2$,
and the error ranges are 95\% confidence bounds on a Poisson distribution.  We
list results in steps of $2\,$mJy.  We also include published results from
other surveys for comparison. These include the comparable results from the
wide area 8-mJy blank field survey, and three separate surveys conducted
toward galaxy clusters, which appear to show a higher estimate than blank
field counts.  }
\scriptsize
\label{tab:hdfsc}
\begin{tabular}{lllll|lll}
\\\hline
$S$[mJy]&$N_{\rm raw}(>{S})$&$\gamma(S)$&$N(>{S})$&$N_{\rm 8\,mJy}(>{S})$&$N_{\rm cluster}(>{S})$&$N_{\rm cluster}(>{S})$&$N_{\rm cluster}(>{S})$\\
&&&&\citep{2002MNRAS.331..817S}&\citep{2002AJ....123.2197C}&\citep{2002MNRAS.331..495S}&\citep{2002MNRAS.330...92C}\\\hline
0.25&   &     &                     &$$                  &                          &$51000_{-21000}^{+21000}$&\\
0.3 &   &     &                     &$$                  &$33000_{-13000}^{+63000}$ &                         &\\
0.5 &   &     &                     &$$                  &$18000_{-9000}^{+12000}$  &$27000_{-10000}^{+10000}$&\\
1   &   &     &                     &$$                  &$9000_{-5000}^{+14000}$   &$9500_{-3400}^{+3400}$   &$15000_{-3900}^{+5700}$\\
2   &19 &3.10 &$1923_{-739}^{+1002}$&$$                  &$3500_{-1000}^{+1500}$    &$2900_{-1000}^{+1000}$   &$6800_{-1900}^{+2600}$\\
3   &   &     &                     &$$                  &                          &                         &$3560_{-1200}^{+1560}$\\
4   &16 &1.36 &$710_{-293}^{+409}$  &$$                  &                          &$1700_{-800}^{+800}$     &$1800_{-690}^{+990}$\\  
5   &   &     &                     &$$                  &$2100_{-300}^{+800}$      &                         &\\
6   &14 &1.07 &$489_{-213}^{+304}$  &$550_{-170}^{+100}$ &                          &$$                       &$1000_{-450}^{+700}$\\  
8   &9  &0.99 &$285_{-149}^{+231}$  &$320_{-100}^{+80}$  &                          &$900_{-580}^{+580}$      &\\       
10  &4  &0.98 &$128_{-90}^{+173}$   &$180_{-60}^{+60}$   &                          &$$                       &$360_{-310}^{+770}$\\
12  &3  &0.98 &$96_{-74}^{+157}$    &$40_{-30}^{+30}$    &                          &$$                       &\\   
14  &2  &0.99 &$65_{-55}^{+142}$    &$$                  &                          &$$                       &\\
16  &1  &1.00 &$33_{-31}^{+121}$    &$$                  &                          &$<420$                   &\\\hline
\normalsize
\end{tabular}
\end{table*}

\subsection{The 450\mum\ number counts}
Since we have only detected 5 objects at 450\mum, we choose to quote only a
single value for $N({>}\,100\,$mJy).  Because the number counts are not well
constrained at all at 450\mum\, it is difficult to know what model to use in
the completeness simulations.  We chose a simple power law form with
$N(S)dS\propto S^{-\alpha}$.  Our estimates of the detectability were largely
insensitive to the choice of $\alpha$, and turned out to be simply the inverse
of the completeness at 100\,mJy (80\%).  This result, along with two previous
estimates at lower fluxes, is given in Table \ref{tab:sc450}.  There are
no published estimates at the bright flux levels which are probed here.

\begin{table}
\caption[450\mum\ source counts from the HDF super-map.]{450\mum\ source
counts from the HDF super-map.  We present here our estimate and those
compiled from the literature.  The other estimates are based on a different
approach and also chose to quote symmetric error bars.}
\label{tab:sc450}
\begin{tabular}{lll}
\\\hline
Flux (mJy) & $N({>S})$ & Comment \\ \hline
10   & $2100^{+1200}_{-1200}$ & \citet{2002MNRAS.331..495S}\\
25   & $500^{+500}_{-500}$    & \citet{2002MNRAS.331..495S}\\
100  & $140^{+160}_{-90}$     & This work \\ \hline
\end{tabular}
\end{table}
A fit to a power law with these combined counts gives $\alpha=2.2\pm0.4$.
This is not too dissimilar from the 850\mum\ slope, though slightly shallower.
This might be an interesting result:  if the SCUBA sources are mainly at
$z\gsim1$, then one would expect the 450\mum\ fluxes to drop off more steeply
than the 850\mum\ counts because the K--correction becomes positive for
450\mum\ at these redshifts.  With only three estimates for the 450\mum\
number counts however, it is premature to draw strong conclusions.
Nevertheless, this shows that constraints on the number counts at different
wavelengths can serve as a probe of the evolution and redshift distribution
of sub-mm galaxies.

\section{Clustering of sub-mm sources}

As we have already mentioned, the presence of clustering may also affect
estimates for the source counts.  \citet{2000MNRAS.318..535P} report
weak evidence of clustering in 
the $\sim2\times2\,$ arcminute map of H98 in
the sense of statistical correlations with LBGs, which are themselves
strongly clustered \citep{2001ApJ...550..177G}.  The UK 8\,mJy survey \citep{2002MNRAS.331..817S}
which covers over 250 square arcminutes, and the smaller yet deeper CFRS
3- and 14-hour fields \citep{webb} fail to detect clustering among the SCUBA detected sources.   There appears (by eye) to be clustering in the
super-map;  in particular the concentration of sources near the centre of the
map, the trio of sources to the west of the map, and the group of 4 north of
it might suggest a clustering scale on the order of 30 arcsec or so.  However,
one would \textsl{expect\/} more sources in these areas because of the
increased sensitivity.  Hence one needs to carry out a statistical
clustering analysis, including the inhomogeneous noise, to quantify this.

Clustering is usually described as the probability, $p$, of finding a source
in a solid angle $\Omega$, and another object in another solid angle $\Omega$
separated by an angle $\theta$.  This probability is described by:
\begin{equation}
\label{equ:ptheta}
p(\theta) = N^2[1+w(\theta)]\Omega^2,
\end{equation}
where $N$ is the mean surface density of objects on the sky and $w(\theta)$ is
the angular two-point correlation function.  If $w(\theta)$ is zero, then the
distribution of sources is completely random, while otherwise it describes the
probability in excess of random.  The correlation function $w(\theta)$ is
estimated by counting pairs and there are several specific estimators in the
literature. The one we employ is that proposed by \citet{1993ApJ...412...64L}:
\begin{eqnarray}
\label{equ:wtheta}
w(\theta) &=& {{DD -2DR + RR}\over{RR}} ,\\
{\rm with}\ \delta w(\theta) &=& \sqrt{{1+w(\theta)}\over{DD}}.
\end{eqnarray}
This particular estimator has been shown to have no bias and also has a lower
variance than the alternatives.  In this equation, $D$ represents sources in
the SCUBA catalogue, and $R$ are sources recovered from Monte-Carlo catalogues.
$DD$ is the number of pairs of real sources that fall within a bin of width
$\delta\theta$ in the map.  $DR$ are data-random pairs, and $RR$ are
random-random.  For simplicity, each catalogue is normalised to have the same
number of objects.  To obtain the random catalogues, we created 1000 mock
fields based on the source count model used in the previous section, and the
actual noise of the real data.  The sources were placed
randomly throughout the field, and the resulting mock data were placed into the same pipeline as our real data. 

This approach is slightly different from that of \citet{webb} and
\citet{2002MNRAS.331..817S}, the only other two surveys to attempt a
clustering analysis of SCUBA sources.  In those analyses, the mock images were
modified only by adding noise to each pixel.  The amount of noise added was
taken from the noise map which was created along with the real signal map.
Therefore their final mock images do not exhibit the chop pattern that one
would expect to see.  Recognizing this limitation, \citet{webb} took the added
step of masking out regions in the mock images that correspond to the positions
of the off-beams in the real map.  Our simulations involve a full sampling of
the mock images using the astrometry information from the real data.
Therefore the simulated and real maps have the same beam features.

\begin{figure}
\begin{center}
\includegraphics[width=3in,angle=0]{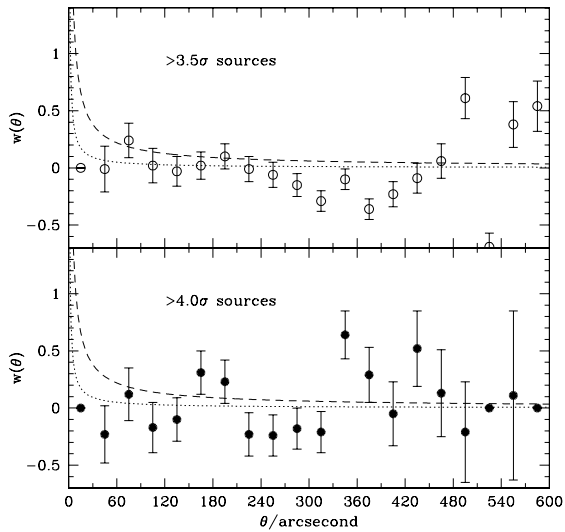}
\caption[Angular two-point correlation function estimate for the HDF super-map
sources.]{Angular two-point correlation function estimate for the HDF
super-map sources. Our estimates of $w(\theta)$ are shown as the open circles
(for the $>3.5\sigma$ catalogue) and filled circles (for the $>4.0\sigma$
catalogue).  The points are plotted at the midpoint of each 30 arcsec bin.  The first point is zero because there are no objects in the super-map closer than 30 arcsec to another.  The measured clustering signal of EROs is shown as a dashed line,
and that of LBGs as a dotted line.  There is no evidence of sub-mm clustering
here, though the errors are still quite large.}
\label{fig:clusterwt}
\end{center}
\end{figure}

We used relatively wide 30 arcsec bins in because the number of sources is so low.   This is also twice the size of the SCUBA beam at 850\mum, which was the binsize criterion used by \citet{2002MNRAS.331..817S}. $w(\theta)$ is estimated using both
the 3.5 and $4\sigma$ catalogues.  Though some of $3.5\sigma$ sources may be
spurious, the increased number of objects helps bring down the clustering
error bars.  As Fig.~\ref{fig:clusterwt} shows, however, even when the
$3.5\sigma$ sources are included there is no evidence for clustering in the
HDF super-map, since there is no angular bin that has a $w(\theta)$
significantly different from zero.  For comparison in that figure, we also
plot estimates of $w(\theta)$ for EROs and LBGs
\citep{2000A&A...361..535D,2001ApJ...550..177G}.  We repeated this using bins half as wide, and again no significant deviation from zero was found.

\begin{figure}
\begin{center}
\includegraphics[width=3in,angle=0]{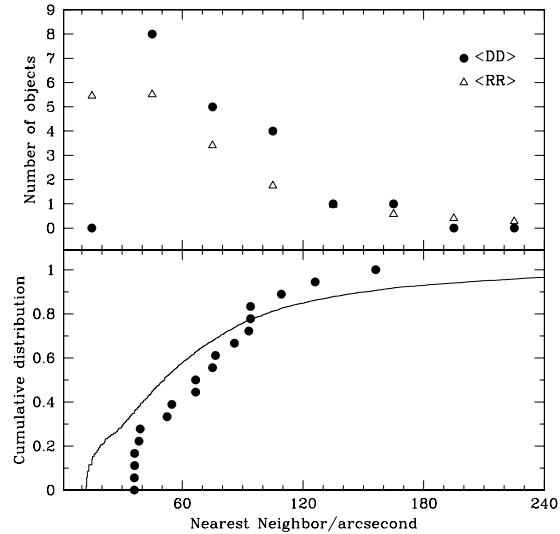}
\caption[Nearest neighbour clustering analysis.]{Nearest neighbour clustering
analysis.  In the top panel we plot the nearest neighbour distribution of the
data compared with a Monte-Carlo generated random set.  The random data falls off below 12 arcseconds, where our source extractor is unable to seperate unique objects. There is a lack of real sources detected with pair separations less than $\sim30\,$arcsec.  This
is also reflected in the bottom panel, where we plot the cumulative
distribution of nearest neighbours (dots) compared with Monte-Carlos
(solid line).}
\label{fig:clusternn}
\end{center}
\end{figure}

This is not the only clustering estimator one can use.  We also performed a
`nearest-neighbour' analysis (see e.g.~Scott \& Tout 1989) to test if sources
were closer together than expected at random.  This statistic simply
examines the distribution of distances to the nearest neighbours for each
source.  The cummulant of nearest-neighbours
is then compared against our set of Monte-Carlo catalogues to determine if a
pair-wise clustering signal is present.  The results are shown in
Fig.~\ref{fig:clusternn}.

There is a lack of sources with neighbors closer than $\sim30$ arcseconds, but a formal Kolmogorov-Smirnov test indicates only a $\sim40\%$ chance that the distributions are significantly different.   So there is no strong evidence of clustering here either. 

Note that in each of these clustering analyses, it is difficult to estimate
the clustering strength on scales near to the beam size.  Our source
extraction algorithm is insensitive to a fainter source closer than 12 arcsec
to a brighter one.  Note that if SCUBA sources are clustered in a similar manner as EROs or LBGs,
then the signal in $w(\theta)$ would be expected to be strongest at
${\lsim}\,30$ arcsec. This is only a factor of 2 larger than the SCUBA beam,
and therefore one expects that a clustering detection with SCUBA will be
rather difficult due to blending of the sources. Our particular approach has
not been optimised for separating nearby sources and hence our catalogue is
not ideal for clustering analysis.   Future attempts to measure the clustering strength should pay particular care to source extraction algorithms.

There is currently no convincing detection of sub-mm clustering; what is 
needed is a very large (${\sim}\,1$ square degree) survey with hundreds of 
sources in
order to decrease the error bars on the clustering estimate.  The on-going
SHADES survey (see \textsl{http://www.roe.ac.uk/ifa/shades/}  ) purports to do just that.

\section{Implications of the counts}
\subsection{The 850\mum\ sub-mm background}
We explored the range of parameters for the two-parameter model that could fit
our source counts and still fall within the limits of the FIR 850\mum\
background constraint: ${\rm 3.1-4.1\times10^4\,mJy\,deg^{-2}}$
\citep{1996A&A...308L...5P,1998ApJ...508...25H,1998ApJ...508..123F}.  By
calculating the integral of $S N(S)dS$ we find that our ${>}\,2\,$mJy sources
contribute $1.4\times10^4\,$mJy$\,{\rm deg}^{-2}$ to the FIB.
This is consistent with estimates from several groups, and demonstrates that
a significant fraction of the sub-mm Universe is still below the flux
threshold attainable from current SCUBA surveys.  However, given the freedom
which still exists in the faint end counts, and in the current level of
uncertainty of the sub-mm background itself, the entire background can
easily be made up of sources with $S_{\rm 850\mu m} >0.1\,$mJy.  This result
can be used to constrain models that predict the evolution of IR luminous
galaxies.  Future surveys that detect more sources will constrain the source
counts further, and extend the limiting flux down to fainter levels.

\subsection{Evolution of sub-mm sources}
The counts are much greater than what one obtains by estimating 850\mum\ fluxes
simply from the \textsl{IRAS} 60\mum\ counts.  As addressed by \citet{blain},
a SCUBA galaxy with a flux $\gsim 5$\,mJy has an inferred luminosity in excess
of $10^{12}{\rm L}_\odot$ if they are distributed at redshifts greater than 0.5.  Note that Chapman et al. (2003) find spectroscopic redshifts $z>0.8$ for each
source in their sample of SCUBA detected radio galaxies.  Such sources make up at least 50\% of the SCUBA population brighter than 5\,mJy, and radio 
\textsl{undetected} sources are likely to be at even higher redshifts.

At these luminosities the number of SCUBA
objects per co-moving volume is several hundred times greater than it is today.
Therefore there must be significant evolution past $z\gsim0.3$ (the
\textsl{IRAS} limiting redshift).  

Modeling this evolution has been difficult due to the lack of observational
data on the redshift distribution of SCUBA sources.  Attempts to model the
luminosity evolution of the SCUBA sources have been carried out using
semi-analytic methods (e.g.~Guiderdoni et al.~1998) and parametric ones
(e.g.~Blain et al.~1999, Rowan-Robinson 2001, Chapman et al. 2003).  In many cases, the starting
point is the well determined \textsl{IRAS} luminosity function.  This gives us
the number of sources of a given luminosity per co-moving volume.  These
luminosities are then modified as a function of redshift, and then 850\mum\
fluxes are extrapolated and source counts determined.   We expect the number
of galaxies to increase in the past due to merger activity, so number
evolution must play some role, but it is noted that strong number evolution
overproduces the FIR background.   There are many models (and a range
of parameters within them) that fit the current data.  Therefore, until we can
better constrain the counts and determine redshifts, the only firm conclusion
one can make is that SCUBA sources do evolve strongly.
Our catalogue of 34 SCUBA sources within the HDF region, with the amount of
multi-wavelength data being collected there, should contribute toward
distinguishing between models.

\subsection{A connection with modern-day elliptical galaxies?}
Observations of local star-forming galaxies suggest that
\begin{equation}
\label{equ:sfr}
{\rm SFR} = 2.1\times10^{-10}\left( L_{\rm 60\mu{m}}/{\rm L}_\odot\right)
 {\rm M}_\odot {\rm yr}^{-1},
\end{equation}
where $ L_{\rm 60\mu{m}} \equiv (\nu S_\nu)|_{60\mu{\rm m}}$ is the 60\mum\
luminosity (Rowan-Robinson et al. 1997).  Assuming the SLUGS result for dust temperature and emissivity (Dunne et al. 2002), we calculate SFRs in excess 
of $1000\,{\rm M}_\odot$yr$^{-1}$ for
redshifts past about 1.  Of course the conversion between detected flux and
inferred star formation rate is highly dependent on the dust SED, and can
change by factors of 10 for changes in temperature and $\beta$ of only 2.
Also, the simple relation between FIR luminosity and SFR may be different for
these more luminous sources \citep{1986ApJ...311...98T,1997MNRAS.289..490R}.

Despite these uncertainties, it has been suggested
(e.g.~Blain et al.~2002, and references therein) that SCUBA sources can be
associated with the elliptical galaxies we see today via the following
argument.  Producing the local massive elliptical population with a
homogeneous stellar distribution requires a sustained period of star formation
on the order of $1000\,{\rm M}_\odot$yr$^{-1}$ lasting about 1\,Gyr.  Based on
results from Chapman et al. (2003) that place the bright SCUBA population at $z>1$,
the number of these galaxies per unit co-moving volume  is comparable to the
density of the local elliptical population.  For example, if we take our
estimate of the counts above $5\,$mJy and assume that they cover a
redshift range between 2 and 4 in a standard flat $\Lambda$-dominated model,
we obtain a density of about $7\times10^{-5}(h^{-1}{\rm Mpc})^{-3}$.
These are thus rare and extremely luminous objects, with comparable
number densities to galaxy clusters or quasars.

If SCUBA sources really are associated with elliptical galaxies, they should
exhibit spatial clustering like their local counterparts.  There are other
reasons one might expect detectable clustering; Extremely Red Objects (EROs)
are very strongly clustered \citep{2000A&A...361..535D}, and seem to have a
correlation with SCUBA sources.  In general objects associated with major
mergers should show high amplitude clustering (e.g.~Percival et al.~2003).

Although our analyses show no sign of clustering, the
data are not powerful enough to rule it out.  To improve on this we
need more detected sources in order to bring down the Poisson error-bars.
Also, the ERO and LBG clustering observations are taken from samples that
exist at a common redshift (${\sim}\,1$ in the case of EROs and ${\sim}\,3$
for LBGs).  Because of the strong negative K--correction, detected SCUBA
sources are spread across a much wider redshift range, therefore diluting the
clustering signal.  Hence, progress can only be made with a larger survey
(such as SHADES) that also has the ability to discriminate redshifts,
even if only crudely.

Although more studies are required to verify this claim, it is
a reasonable hypothesis, and one with some testable predictions. Our new
catalogue of SCUBA sources (Table~2) should allow for future detailed
comparison with other wavelength data, which facilitate such tests.

\section{Conclusions}
This paper has presented the most complete accounting of sub-mm flux in the
HDF-North region to date.  We were able to demonstrates that ${>}\,4\sigma$
SCUBA detections are quite robust, being
consistently detected in independent observations of the same area.
Our catalogue of sources was obtained using a careful statistical approach,
involving simulations with the same noise properties as the real data.  At
850\mum\ we were able to extract 19 sources above $4\sigma$ and a further
15 likely sources above $3.5\sigma$.  Such a large list, in a field with
so much multi-wavelength data, should be extremely useful for further studies.

Our estimated source counts cover a wider flux density range than any other
estimates, and given the careful completeness tests we carried out, they
are likely to be more reliable than combining counts from different surveys.
Our counts of SCUBA sources verify that
significant evolution of the local LIRG population is required.
Extrapolating a fit to these counts to below $2\,$mJy can
reasonably recover the entire FIB at 850\mum.  Clustering, although
anticipated to be strong, was not detected in our map, due largely to the
limited number of sources.  Several hundred sub-mm sources with at least some
redshift constraint will be required to detect the clustering unambiguously.

The power of the SCUBA observations in the HDF-N lies not in the detection of
objects per se, but rather for the ability to compare them
with the plethora of existing and upcoming deep maps of this region at a
wide variety of other wavelengths.  Some of these comparisons will be
the focus of paper II.

\section*{Acknowledgments}

This work was supported by the Natural Sciences and Engineering Research
Council of Canada. The James Clerk Maxwell Telescope is operated by The
Joint Astronomy Center on behalf of the Particle Physics and Astronomy
Research Council of the United Kingdom, the Netherlands Organisation
for Scientific Research, and the National Research Council of Canada.
Much of the data for this paper was obtained via the Canadian Astronomy Data Centre, which is operated by the Herzberg Institute of Astrophysics, National Research Council of Canada.  We also thank Amy Barger for access too some of her data prior to their release in the public archive.

\end{document}